# Magnetotransport and Carrier Dynamics in Quasi-One-Dimensional Antiferromagnet KMn$_6$Bi$_5$

Qi-Yuan Liu[1,2], Chenfei Shi[1], Zhaodi Lin[1], Xiutong Deng[2], Baojuan Kang[1], Youguo Shi[2], Gang Wang[2], Rongrong Jia[1*], and Jin-Ke Bao[3*]

1Department of Physics, Materials Genome Institute, Institute for Quantum Science and Technology Shanghai University, Shanghai 200444, China
[2]Beijing National Laboratory for Condensed Matter Physics, Institute of Physics, Chinese Academy of Sciences, Beijing 100190, China
[3] School of Physics and Hangzhou Key Laboratory of Quantum Matters, Hangzhou Normal University, Hangzhou 311121, China

*Corresponding authors. Email: r.jia@t.shu.edu.cn; jinke_bao@hznu.edu.cn

Quasi-one-dimensional materials $A$Mn$_6$Bi$_5$ ($A$ = Na, K, Rb, Cs) exhibit unique electronic behaviors such as antiferromagnetism, charge density waves, and pressure-induced superconductivity. Thus, they serve as a suitable model system to investigate emergent quantum phenomena produced by the interactions among spin, charge, and lattice. Here we report the magnetotransport properties of KMn$_6$Bi$_5$, revealing a cascade of temperature-dependent carrier dynamics. Below 5 K, the system, despite its anisotropic electronic structure, could be effectively described by an isotropic two-band model and exhibits a large, non-saturating magnetoresistance ($\propto B^{1.8}$). Upon warming, a crossover to a single-band regime occurs around 20 K, driven by the suppression of a hole pocket. Electron density recovers as antiferromagnetic gap openings gradually close from 25 to 70 K which is just below the Néel temperature. Within this temperature range, field-quenched spin fluctuations suppress magnetoresistance. Furthermore, we attribute the low-temperature resistivity upturn to the scaling behavior of magnetoresistance. These findings provide crucial insights into the interplay of dimensionality, magnetism, and electron correlations in quasi-one-dimensional magnetic semimetals.

***Introduction.*** In contrast to three-dimensional (3D) systems, electrons confined to one dimension exhibit unique phenomena such as highly correlated collective excitations, spin-charge separation (decoupling), and nonlocal order parameters due to enhanced interactions.[1-3] Consequently, Tomonaga-Luttinger liquid (TLL) theory, rather than Fermi liquid theory, provides the essential framework for describing these systems.[4] They have garnered significant interest over recent decades, which stems not only from the availability of simple, solvable theoretical models enabling direct theory-experiment comparisons but also from their potential to host novel quantum phenomena, such as Majorana zero modes.[5-7]

While most real materials are 3D, they can exhibit properties similar to ideal one-dimensional (1D) systems under conditions of extreme uniaxial anisotropy. Examples include nanowires with cross-sectional dimensions much smaller than their length. Even though these materials are macroscopically 3D, they can also exhibit 1D properties due to 1D active building blocks with weak coupling between neighboring blocks. Such systems are classified as quasi-one-dimensional (Q-1D) materials and host rich emergent quantum phenomena. Charge density waves (CDWs), a manifestation of the Peierls instability, have been first observed in organic Q-1D Bechgaard salts like (TMTSF)$_2$PF$_6$, which also show spin density waves (SDWs) and pressure-induced superconductivity.[8,9] Femtosecond spectroscopy in Q-1D $K_{0.3}MoO_3$ reveals collective CDW modes and single-particle excitations across the Peierls gap.[10] The spin-Peierls transition, a dimerization instability of antiferromagnetic (AFM) chains, is observed in inorganic Q-1D $CuGeO_3$, evidenced by a collapse in the magnetic susceptibility below 14 K.[11] Unconventional superconductivity emerges in Q-1D $K_2Cr_3As_3$ ($T_c \approx 6$ K),[12] with the estimated $H_{c2}$ exceeding the Pauli limit by a factor of 3–4. $SrCuO_2$ realizes nearly ideal $S=1/2$ Heisenberg AFM chains, demonstrated by logarithmic corrections to the susceptibility at low temperatures.[13] These phenomena underscore the role of reduced dimensionality in stabilizing novel quantum phases, although real materials inevitably exhibit interchain coupling that modifies ideal 1D physics.[14,15]

Among the diverse landscape of Q-1D compounds, transition-metal-based systems are particularly intriguing due to the complex interplay between magnetic order, electronic correlations, and potential superconductivity.[16,17] A prime example is the $A$Mn$_6$Bi$_5$ family ($A$ = Na, K, Rb, Cs). These compounds crystallize in a monoclinic structure (space group $C2/m$) and feature double-walled [Mn$_6$Bi$_5$]$^-$ columns extending along the $b$-axis, separated by alkali metal ions.[18-21] They exhibit AFM order below Néel temperatures $T_N \approx$ 81 K (Cs), 83 K (Rb), 75 K (K), 47–85 K (Na).[18-22] Notably, pressure suppresses the AFM order in KMn$_6$Bi$_5$,[23] RbMn$_6$Bi$_5$,[24] and CsMn$_6$Bi$_5$,[21] inducing bulk superconductivity with $T_c$ up to 9.5 K near quantum critical points. The resulting superconducting dome adjacent to the AFM phase, along with $H_{c2}$ exceeding the Pauli paramagnetic limit near the critical pressure $P_c$, strongly suggests unconventional pairing mediated by magnetic fluctuations, drawing parallels to heavy-fermion and iron-based superconductors.[25,26] This family also displays intriguing diversity: CDWs and SDWs coexist in KMn$_6$Bi$_5$ crystals, RbMn$_6$Bi$_5$ is predicted to host a helical spin state,[20] while NaMn$_6$Bi$_5$ undergoes complex magnetic transitions and notably lacks pressure-induced bulk superconductivity.[22] A recent study on this family of compounds shows that the charge carrier type can be modulated by the radius of alkali metal ions, as determined from Seebeck coefficient and Hall resistance measurements.[27]

At ambient pressure, KMn$_6$Bi$_5$ exhibits pronounced electronic anisotropy, with a resistivity ratio $\rho_\perp/\rho_\parallel \approx 25$ at low temperatures[18,28] ($\rho_\perp$ denotes resistivity perpendicular to the $b$-axis, while $\rho_\parallel$ represents resistivity along the $b$-axis). This reflects its uniaxial structure: metallic conduction is favored along the $b$-axis within the Mn-cluster core columns, while intercolumn electron tunneling shows an incoherent-to-coherent crossover below ~40 K. The AFM transition at 75 K is characterized by a small effective moment (1.56 $\mu B$/Mn, as determined from Curie-Weiss fits in the paramagnetic state), exhibiting SDWs.[18]

Despite significant progress in understanding its basic properties,[18,23,28,29] the magnetic response of charge transport within or between the [Mn$_6$Bi$_5$]$^-$ columns demands further investigation. Moreover, a $\rho_\parallel(T)$ upturn resembling a field-induced metal-insulator transition[30-47] has been observed at low temperatures in several typical semimetal families but such behavior in this Q-1D family remains unexplored.

In this work, we investigate the magnetotransport behavior of KMn$_6$Bi$_5$ single crystals through comprehensive magnetoresistance (MR) and Hall measurements. We identify a phenomenological electron-hole compensation below 5 K, a crossover from two-band to single-band transport around 20 K, and show that the low-temperature $\rho_\parallel(T)$ upturn is a natural consequence of MR scaling with $\gamma > 1$,[48] rather than a phase transition. These results clarify key aspects of carrier dynamics in this Q-1D magnetic semimetal.

***Crystal growth and physical property measurements***. High-quality single crystals of KMn$_6$Bi$_5$ were grown using a self-flux method as previously reported.[18] Electrical transport properties were measured using a standard four-probe geometry in a Quantum Design PPMS over 1.8–300 K and magnetic fields up to ±14 T. MR and Hall resistivity were simultaneously acquired using a five-probe configuration. All the measurement preparation was performed under inert conditions due to the crystals' air sensitivity. To ensure reproducibility, data from four different batches were collected, with the main analysis based on sample A. Sample handling and measurement details are provided in the *Supplemental Information*.

***Resistivity***. Fig. 1(a) shows $\rho_{xx}$ (denoted as $\rho_\parallel$ previously for consistency) of KMn$_6$Bi$_5$

sample B where $\rho_{xx}$ stands for the temperature-dependent resistivity with the current applied parallel to b-axis. Fields of 12 T and 0 T were applied perpendicular to the rod axis. The history of the applied field and temperature variation has little effect on the longitudinal resistivity, $\rho_{xx}(T)$. A small kink at 75 K corresponds to the AFM transition. The overall shape of the curve and the AFM transition temperature are identical to those in previous reports.[18] However, a reproducible metal-insulator-like upturn is observed under a high magnetic field below 10 K. This phenomenon will be addressed later after analyzing how MR(B) governs $\rho(B,T)$. The first-order derivative of $\rho_{xx}(B=0,T)$ reaches a peak around 20 K, indicating the rapid change in resistivity which will be discussed later.

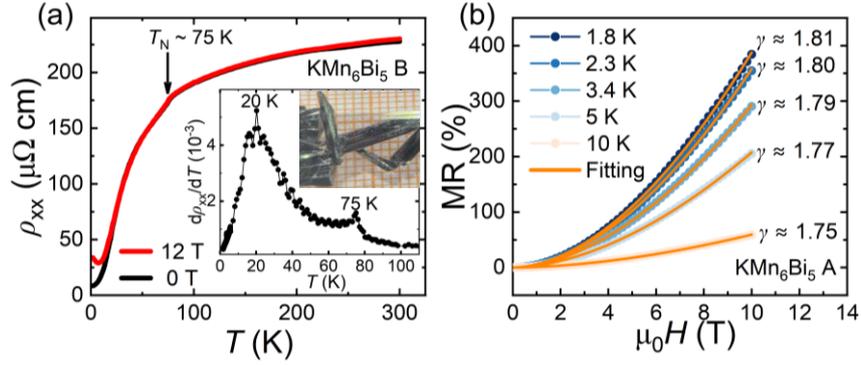

**Fig. 1.** (a) Temperature-dependent resistivity $\rho_{xx}$ of KMn$_6$Bi$_5$ sample B measured at 0 and 12 T over 1.8–300 K ($I \parallel rod$, $H \perp I$). Inset shows the first derivative of the longitudinal resistivity versus temperature ($d\rho_{xx}/dT$) at 0 T in the range 1.8–110 K. Temperatures corresponding to the two characteristic peaks are marked. Large, high-quality single crystals of KMn$_6$Bi$_5$ on 1 mm grid lines are also shown. (b) MR versus magnetic field for sample A at four selected temperatures below 10 K (0–10 T). Orange curves show fits to Kohler's scaling (MR $\propto (B\tau)^\gamma$). Slight deviation above 10 T arises from field-dependent reduction in $\gamma$, necessitating segmented fitting with variable $\gamma$ values.

*Magnetoresistance.* Kohler's rule is typically applicable only to single-band systems or two-band systems with full electron-hole compensation.[49] A unified parameter $B\tau$ establishes a robust scaling framework for magnetotransport analysis, provided that $\tau$ is the sole temperature-dependent scattering parameter. In KMn$_6$Bi$_5$, phenomenological near-electron-hole compensation below 5 K—accompanied by matched temperature dependencies of densities and mobilities for both charge carriers—closely approaches the above requirements, enabling an approximate description via $B\tau$-governed MR power laws. Although minor discrepancies in scaled MR curves reveal violations of Kohler's rule (attributed to $\tau_e/\tau_h \neq \text{Const.}$), the $B\tau$ scaling retains validity in interpreting complex $\rho(B,T)$ behavior according to Ref.[48] The scaling behavior of MR can be expressed as follows: $\text{MR} = \frac{\rho(B)-\rho_0}{\rho_0} \propto (B\tau)^\gamma = \left(\frac{B}{\alpha\rho_0}\right)^\gamma$ (1). The longitudinal resistivity is expressed as: $\rho(B,T) = \rho_0(T) + \zeta \frac{B^\gamma}{(\alpha\rho_0)^{\gamma-1}}$ (2). Detailed discussions on the above equations are covered in Ref.[48]

Fig. 2 shows the MR of KMn$_6$Bi$_5$ sample A as a function of magnetic field at different temperatures using the same configuration as for $\rho_{xx}$ measurements ($I \parallel rod$, $H \perp I$). According to the relation MR $= [\rho(H) - \rho(0)]/\rho(0) \times 100\%$, MR reaches 647% at 1.8 K and 14 T, exhibiting no sign of saturation. The MR primarily manifests a non-saturating behavior with approximately $H^2$ dependence, which is similar to many known materials exhibiting XMR, such as WTe$_2$, Cd$_3$As$_2$, and TaAs.[30,50,51]

To further investigate the non-saturating MR, we use equation (1) to fit MR below 10 K, as shown in Fig. 1(b), applying Kohler's rule. The exponent $\gamma$ is related to the degree of carrier compensation, and for systems exhibiting perfect electron-hole compensation, $\gamma = 2$.[52,53] We find $\gamma$ is approximately 1.8, with a slight decline (from 1.81 to 1.77) as temperature rises from 1.8 K to 5 K ($R^2 > 99.98\%$). Concurrently, MR (14 T) drops from 647% to 345%.

Typically, the magnetic response of carriers depends on carrier type. For a single carrier type, the Lorentz force causes transverse deflection, establishing a Hall voltage that, under ideal conditions, balances the Lorentz force, resulting in zero MR. In contrast, for nearly compensated electrons and holes, their opposite charges and velocities cause the Lorentz force to deflect both types in the same transverse direction. This co-accumulation hinders a stable Hall voltage due to charge neutralization, preventing force balance. Carriers thus experience sustained deflection, curving their trajectories and prolonging the carrier path length, leading to large, non-saturating MR in the current direction, even with imperfect compensation. The observed $\gamma$ and MR values suggest the presence of phenomenological electron-hole compensation in KMn$_6$Bi$_5$ single crystals below 5 K.

It is important to clarify the term "compensation" used in this context. Strictly speaking, a system is compensated only if the electron and hole densities are equal ($n_e \approx n_h$). As discussed in Hall effect analysis later, our two-band model fitting results below 5 K yield $n_h \approx 2.5 n_e$, indicating that KMn$_6$Bi$_5$ is technically an uncompensated semimetal. However, semiclassical theory predicts that while any density imbalance ($n_e \neq n_h$) will eventually lead to MR saturation at sufficiently high fields, this saturation field can be impractically large for high-mobility carriers. In KMn$_6$Bi$_5$, we observe a large, non-saturating MR that follows a nearly quadratic $B^{1.8}$ power law up to our maximum field of 14 T. This behavior is phenomenologically indistinguishable from that of a truly compensated system within our experimental window. Therefore, in this paper, the term "compensation" is used to describe this observed transport behavior, which is dominated by two-carrier effects, rather than to imply a strict equality of carrier densities.

Within the temperature range 5-17 K, the rapid decline in MR is attributed to a decrease in carrier mobility, accompanied by the gradual deterioration of electron-hole compensation. This occurs because holes, with large effective mass, are more sensitive to phonon scattering than electrons. In the temperature range of 17-25 K, MR at 14 T decreases from 15.7% to 1.3%. The MR vanishing point near 20 K, corresponding to the peak in $d\rho_{xx}/dT$ discussed earlier, signals a gradual transition from two-band to single-band, electron-dominated behavior. During this process, since $\mu_h$ drops faster than $\mu_e$, the contribution of holes to transport fades, resulting in an increase in $\rho_{xx}$, breaking the electron-hole compensation, and leading to the suppression of MR. Hall effect analysis confirms this mechanism. Another possible mechanism for this process is detailed in Hall analysis.

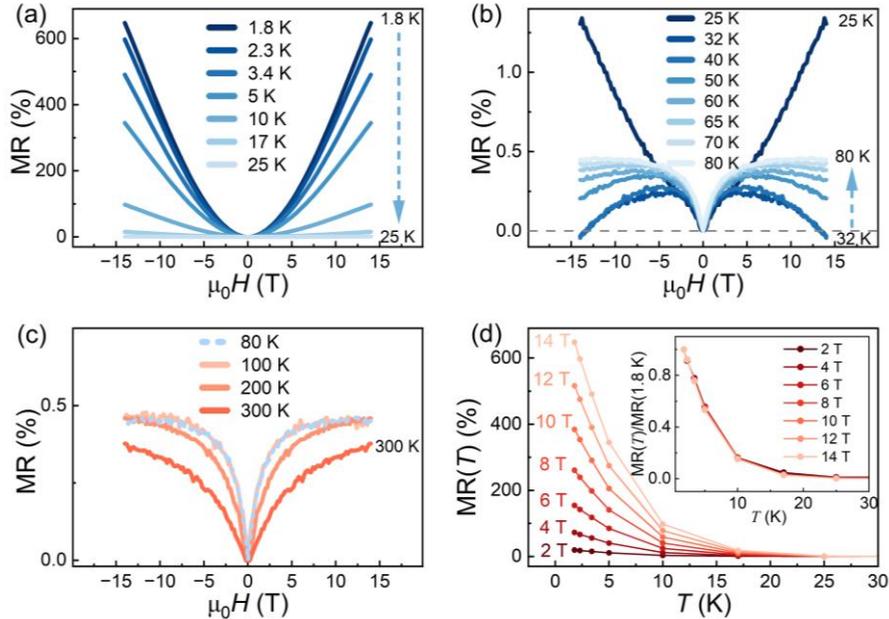

**Fig. 2.** MR measurements and analysis. (a)-(c) MR versus magnetic field (0–14 T) for KMn$_6$Bi$_5$ sample A at selected temperatures from 1.8 to 300 K, measured with the configuration identical to that in Fig. 1(a). (d) Temperature dependence of MR at fixed magnetic fields (2–14 T) derived from field-sweep measurements (1.8–25 K). Inset:

normalized MR curves demonstrating identical temperature dependence.

As shown in Fig. 2(b), once MR is suppressed and the system is dominated by one type of carrier above 25 K, KMn$_6$Bi$_5$ exhibits a complex and distinct MR response in its AFM state ($T$ < 75 K). At 32–65 K, MR initially increases with applied magnetic field ($B \perp I$), saturating below 0.5% for fields of 4–6 T. Upon increasing the field beyond 6 T, MR decreases gradually. Notably, at 32–40 K, MR diminishes to zero near 13 T and transitions to -0.04% at 14 T. According to later discussion of the Hall effect, the carrier system is n-type with relatively low mobility (0.012–0.00059 m²·V$^{-1}$·s$^{-1}$) at 32−60 K.

The exceptionally small magnitude of the saturated MR (< 0.5%) is attributed primarily to the low carrier mobility, where the criterion $\mu B \ll 1$ is almost satisfied ($\mu$ denoting mobility and $\mu B$ < 0.16). This condition strongly suppresses the classical orbital MR, limiting the Lorentz-force-induced modification of electron trajectories. At 25 K, $\mu B$ < 0.35, so the criterion is not satisfied. MR (25 K) reaches 1.3% at 14 T but no saturation is observed.

The gradual decrease of MR at fields exceeding 6 T, culminating in the observed sign change near 13–14 T at 32–40 K, is most consistently explained by field-induced suppression of AFM spin fluctuations. The high magnetic field progressively polarizes the spin system and reduces the amplitude of low-energy spin fluctuations. This suppression leads to decreased fluctuation-mediated electron scattering, reducing resistivity and manifesting as negative MR. The vanishing MR at ~13 T and subsequent small negative MR at 14 T may signify approaching or crossing a field-driven critical point. The quenched spin fluctuation scattering at 32–40 K compared to higher temperatures suggests weaker spin fluctuations in this lower temperature regime within the AFM state. As temperature increases (50–70 K), thermally activated AFM spin fluctuations become stronger and can no longer be quenched at 13 T. However, the downward trend of the MR curves implies that these fluctuations could be quenched at higher magnetic fields.

Above the AFM transition temperature ($T$ > 75 K), the MR is small (< 0.5%) and saturates for fields exceeding 6 T. Crucially, the MR profile remains flat at high fields without exhibiting any downturn. This absence of field-induced resistance suppression confirms the lack of coherent AFM fluctuations in the paramagnetic phase. The observed saturation and small magnitude are consistent with the low carrier mobility. The resulting transport falls into the low-field limit ($\mu B$ < 0.005), where orbital magnetoresistance is negligible.

***Origin of $\rho(T)$ upturn.*** The metal-insulator-like upturn in $\rho_{xx}(T)$ of the 12 T curve (red) at low temperatures in Fig. 1(a)—is not indicative of a phase transition but arises intrinsically from Kohler's scaling behavior of MR in equation (1). This conclusion is supported by the normalized temperature-dependent MR curves in different magnetic fields [inset, Fig. 2(d)]. The collapse of all curves onto a universal function demonstrates identical temperature dependence, thereby excluding a field-induced metal-insulator transition.[54]

Below 10 K, MR follows the scaling form MR $\propto \left(\frac{B}{\alpha \rho_0}\right)^\gamma$ with $\gamma \approx 1.8$. Consequently, the second term in Eq. (2), $\zeta \frac{B^\gamma}{(\alpha \rho_0)^{\gamma-1}}$, decreases with increasing $T$ (at fixed $B$) due to the growth of $\rho_0(T)$. Competition between the first term $\rho_0(T)$ and this field-dependent term explains the observed upturn: At high fields (e.g., 12 T), the enhanced $B^\gamma$ amplifies the second term, dominating $\rho(B,T)$ and yielding a decreasing trend with temperature. Thus, the upturn reflects that MR is governed by both magnetic field and temperature through a power law dependence, rather than electronic phase instability.[48] Conversely, at low fields (e.g., 4 T), $\rho(B,T)$ is primarily governed by $\rho_0(T)$'s monotonic increase with temperature, consistent with the absence of an upturn in Fig. 3(a).

***Crystal quality effects.*** Apart from magnetic field effects, crystal quality also modulates the second term in Eq. (2), $\zeta \frac{B^\gamma}{(\alpha \rho_0)^{\gamma-1}}$. While $\zeta$ is a material-specific constant, the relaxation time $\tau$—inversely proportional to $\alpha \rho_0$—is reduced by lattice displacements and impurities in lower-quality crystals. This suppression diminishes the field-dependent term, causing $\rho(B,T)$ to be dominated by the first term $\rho_0(T)$. Consequently, the high-field upturn vanishes, as experimentally confirmed for sample D at 12 T [Fig. 3(b)]; we attribute this to its inferior crystallinity (discussed later).

Crystal quality further governs the magnitude of MR. The residual resistivity ratio RRR ≡ $\rho(300\ K)/\rho(1.8\ K)$ quantifies sample purity: RRR values of 26 (A), 25 (C), and 21 (D) establish sample A as the highest-quality crystal. Correspondingly, MR at 1.8 K reaches 647% (A), 489% (C), and 281% (D) under 14 T [Fig. 3(c)]. Enhanced purity elongates $\tau$, extending carrier mean-free paths and amplifying Lorentz-force-driven MR. The saturation tendency of MR may also relate to crystal quality. Defects (e.g., vacancies) in crystals may shift the chemical potential $E_F$, undermining electron-hole compensation, and yields a saturation tendency of MR in lower fields. The decay of $d(MR)/dH$ [Fig. 3(d)] clearly shows saturation occurs earlier in lower-RRR samples. Similar observations on crystal quality and MR are reported in systems with electron-hole compensation like $CaBi_2$[46] and NbP[55,56] where superior crystals exhibit larger, non-saturating MR.

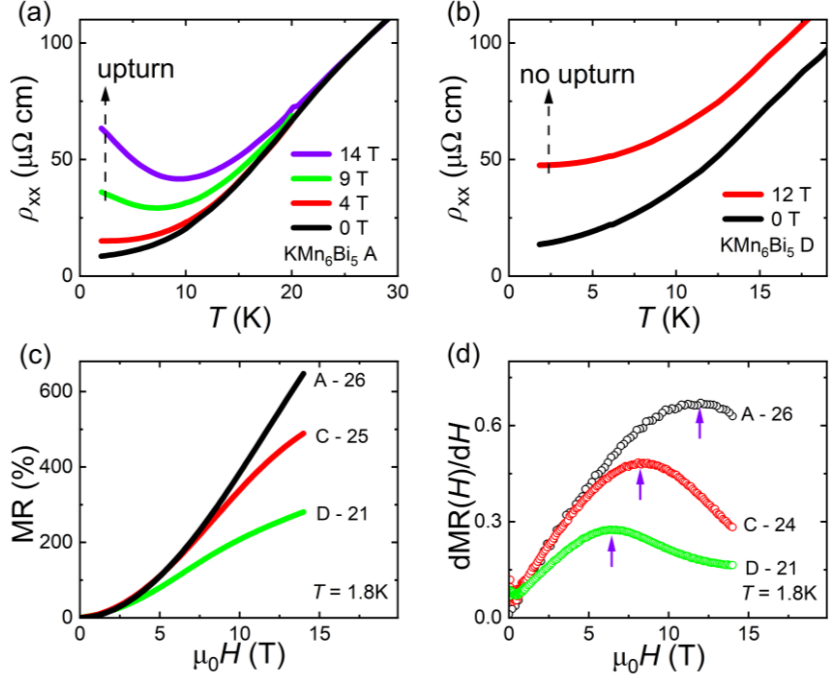

**Fig. 3.** Combined effects of magnetic field and crystal quality on resistivity and MR. (a) $\rho_{xx}(T)$ of $KMn_6Bi_5$ sample A in selected magnetic fields (0–14 T), 1.8–30 K. The arrow marks emergence of a metal-insulator-like upturn at higher fields. (b) $\rho_{xx}(T)$ for sample D at 0 T and 12 T (1.8–18 K). The arrow shows the absence of low-temperature upturn at 12 T. (c) MR versus magnetic field (0–14 T) at 1.8 K for samples D, C, and A. Residual resistivity ratio (RRR) values annotated. (d) Field-derivative of MR [$d(MR)/dH$] from (c). Arrows mark peak positions where $d(MR)/dH$ maxima indicate sign of MR saturation tendency.

*Hall effect.* Fig. 4 shows Hall effect measurements for $KMn_6Bi_5$ sample A. The configuration of applied current and magnetic field is shown in the schematic in Fig. 4(a). Below 7 T, Hall resistivity $\rho_{yx}$ exhibits an approximately linear relationship with negative slope. Above 7 T, $\rho_{yx}$ curves upward at temperatures below 10 K. Curvature increases with decreasing temperature. At 1.8 K, the high-field region exhibits a positive slope. The nonlinear behavior of $\rho_{yx}$ indicates a two-band transport regime below 10 K. In contrast, the near-linearity above 25 K supports using a single-band model. Thus, different models should be applied to different temperature regimes.

To explain nonlinear $\rho_{yx}$ at low temperatures, we employ a two-band model. The well-known expressions relate carrier densities and mobilities to $\rho_{xx}$ and $\rho_{yx}$:

$$\rho_{xx} = \frac{1}{e}\frac{(n_e\mu_e+n_h\mu_h)+(n_e\mu_h+n_h\mu_e)\mu_e\mu_h B^2}{(n_e\mu_e+n_h\mu_h)^2+(n_h-n_e)^2\mu_e^2\mu_h^2 B^2},$$
$$\rho_{yx} = \frac{B}{e}\frac{(n_h\mu_h^2-n_e\mu_e^2)+(n_h-n_e)\mu_e^2\mu_h^2 B^2}{(n_e\mu_e+n_h\mu_h)^2+(n_h-n_e)^2\mu_e^2\mu_h^2 B^2}, \qquad (3)$$

where $n_e$ and $\mu_e$ represent electron density and mobility, and $n_h$ and $\mu_h$ denote hole density and mobility. Fitting results are displayed in Fig. 5, with coefficients of determination $R^2$ exceeding

99.5% and all parameters fall within reasonable ranges and vary with temperature.

Below 5 K, transport is predominantly electronic, as the advantage in electron mobility ($\mu_e \approx 20\,\mu_h$) outweighs the disadvantage in concentration ($n_h \approx 2.5\,n_e$). As theoretically expected, carrier mobility decreases with rising temperature due to a shortened relaxation time $\tau$. Fitting results confirm this trend (1.8 K → 5 K): $\mu_e$ decreases from 0.656 m²·V⁻¹·s⁻¹ to 0.416 m²·V⁻¹·s⁻¹, while $\mu_h$ declines from 0.031 m²·V⁻¹·s⁻¹ to 0.022 m²·V⁻¹·s⁻¹. Carrier density changes are smaller than mobility changes: $n_e$ increases from $1.01\times10^{26}$ m⁻³ to $1.13\times10^{26}$ m⁻³, while $n_h$ decreases from $2.67\times10^{26}$ m⁻³ to $2.41\times10^{26}$ m⁻³. Ratios $n_e/n_h$ and $n_e\mu_e^2/n_h\mu_h^2$ remain basically unchanged, explaining the stability of both $\gamma$ and phenomenological electron-hole compensated behavior, consistent with previous MR analysis below 5 K.

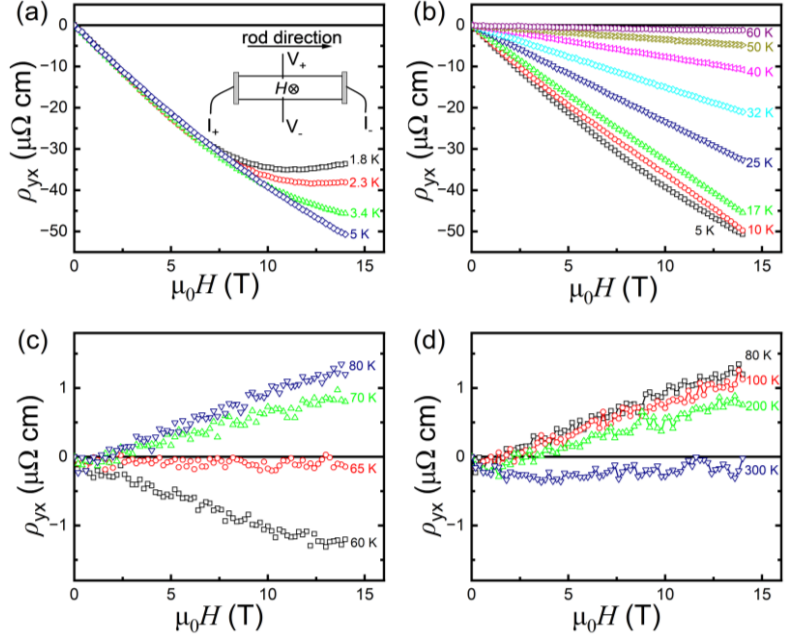

**Fig. 4.** Hall effect measurements. (a)-(d) Magnetic field dependence of Hall resistivity $\rho_{yx}$ for KMn₆Bi₅ sample A at selected temperatures (1.8–300 K, 0–14 T). The schematic in (a) shows the configuration of contacts, current, and magnetic field.

As temperature increases above 5 K, the hole mobility $\mu_h$ decreases drastically. At 10 K, two-band fitting yields $n_h:n_e \approx 5:1$ and $\mu_h:\mu_e \approx 1:36$. Enhanced phonon scattering reduces carrier relaxation times, causing pronounced reductions in both $\mu_h$ and $\mu_e$. Holes, with greater effective mass, are more sensitive to phonon scattering and therefore $\mu_h$ declines relatively faster (73% vs. 49% for $\mu_e$ from 5 to 10 K). Near 17 K, $\rho_{yx}$ exhibits an approximate linear response in magnetic field, signaling a transition to single-band behavior as $\mu_h$ plummets (0.006 → 0.00011 m²·V⁻¹·s⁻¹) due to enhanced scattering. Note that as $\mu_h$ becomes quite small at 10 and 17 K, the overall magnetotransport behavior loses sensitivity to $n_h$, suggesting that the fitted increase in $n_h$ is not physically significant but is likely a fitting artifact that compensates for the sharp decline in $\mu_h$ or the inherent simplifications of the model. At 17-25 K, the vanishing $\mu_h$ eliminates contributions of holes to transport. Concurrently, $\mu_e$ drops to 32% of its initial value while $n_e$ stabilizes, accounting for the $d\rho_{xx}(T)/dT$ peak in this electron-dominated temperature regime.

While KMn₆Bi₅ exhibits pronounced Q-1D characteristics, evidenced by its strong resistivity anisotropy ($\rho_\perp/\rho_\parallel \approx 20$ at 2 K) and near-unity magnetic susceptibility ratio ($\chi_\perp/\chi_\parallel \approx 0.9$),[18] its magnetotransport properties are exceptionally well-described by a simple, isotropic two-band model, with a gradual degraded fitting fidelity across temperature regimes ($R^2 > 99.9\%$ below 3.4 K, $\approx$ 99.6% at 5 K, $\approx$99.4% at 10 K, $\approx$99.0% at 17 K) for b-axis ($I \parallel b, B \perp I$).

To account for this emergent quasi-isotropy, we propose a model based on the low-temperature coexistence of a primary, Q-1D electron pocket and a secondary, more 3D hole pocket. We posit that strong electronic correlations within this two-carrier system renormalize the effective masses, leading to a collective response that can be effectively described by isotropic equations.

This delicate, correlated balance is disrupted in the 10-25 K range as the hole pocket's contribution to conduction is progressively suppressed. This suppression, likely driven by an interplay of dramatic scattering-suppression (evidenced by the sharp reduction in $\mu_h$) and a possible thermal depopulation of hole pocket, provides a mechanism for all associated anomalies. The removal of the hole pocket simultaneously breaks the balance required for the emergent quasi-isotropy, drives the system from a two-band (non-linear Hall) to a single-band (linear Hall) state, and quenches the large MR by destroying the phenomenological compensation condition. The smooth nature of this crossover is consistent with thermally activated scattering.

This physical picture, incorporating a carrier hierarchy, correlations, and band structure evolution, thus offers a self-consistent framework for the observed phenomena. Definitive proof and disentanglement of the underlying mechanisms, however, require direct probes of the Fermi surface, such as angle-resolved photoemission spectroscopy (ARPES).

In the temperature range 25-300 K, a single-band model is adopted to describe $\rho_{xx}$ and $\rho_{yx}$. Corresponding fitting results are shown in Fig. 5(c)(d) and Fig. 6, where $R_H$ denotes the Hall coefficient. The carrier mobility $\mu$ exhibits the expected decreasing trend with temperature, while the carrier density $n$ increases rapidly. An anomaly occurs near 65 K, where $n$ peaks at $6.2 \times 10^{28}$ m$^{-3}$ and $\mu$ reaches a minimum at $6 \times 10^{-5}$ m$^2 \cdot$V$^{-1} \cdot$s$^{-1}$. This anomaly may not be intrinsic and likely originates from the failure of the single-band model in this regime: near-zero $|\rho_{yx}|$ [Fig. 4(c)] amplifies errors in $n \propto |1/\text{slope}|$. A two-band model with $n_e \approx n_h$ and $\mu_e \approx \mu_h$ could explain the small $|\rho_{yx}|$ exhibited in the transition process from electron dominated to hole dominated. Although the system may be compensated, the low mobility greatly suppresses MR, resulting in the absence of recognizable abnormal behavior in MR. Band structure evolution likely underpins this transition, warranting theoretical or ARPES studies.

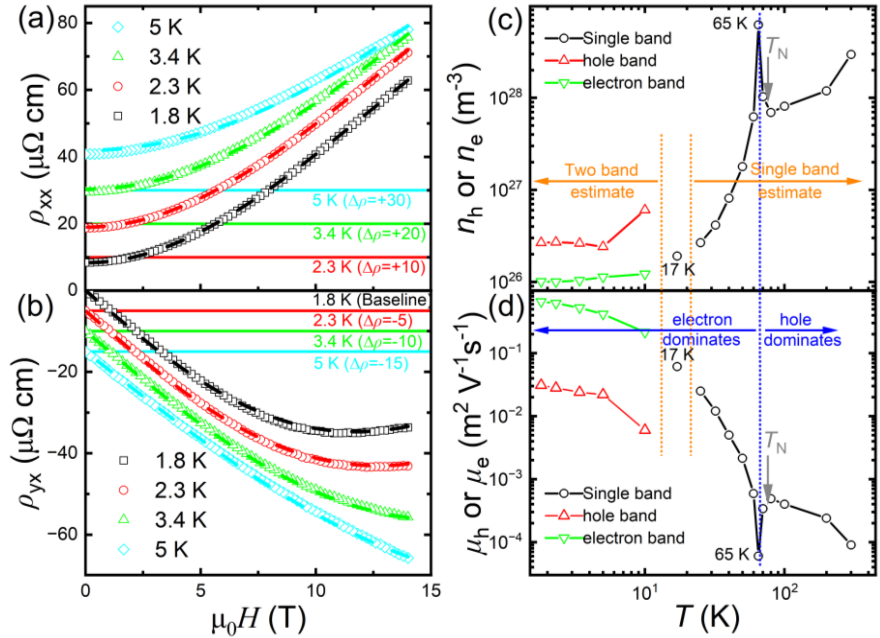

**Fig. 5.** Fitting results based on two-band and single-band models. (a) Longitudinal resistivity $\rho_{xx}$ of KMn$_6$Bi$_5$ sample A at four temperatures below 5 K (0–14 T). Color-matched dashed curves show two-band fits to $\rho_{xx}$. Curves are vertically offset by +10 μΩ cm for clarity; colored baselines indicate original positions. (b) Corresponding Hall resistivity $\rho_{yx}$ with two-band fits (dashed). Curves offset by -5 μΩ cm; baselines marked. (c) Carrier concentrations and (d) mobilities derived from two-band and single-band models (1.8–300 K). Logarithmic scales used throughout. Arrow annotations: Orange: Model-valid temperature regimes; Blue: Electron- or hole-dominated transport regions; Gray: the AFM transition at 75 K

Below 60 K, negative $R_H$ confirms electron-dominated transport. The rapid decrease in $n$ from

60 to 25 K stems from Fermi surface reconstruction due to AFM order, which partially gaps electron bands and reduces carrier density. This process drives sharp reductions in $n$ below $T_N$ = 75 K [Fig. 5(c)]. Over the same temperature range, decreased scattering from AFM spin fluctuations and phonons upon cooling leads to a sharp increase in $\mu$ below $T_N$. These abrupt changes in $n$ and $\mu$ contrast with the gradual variations observed above $T_N$. The $R_H$-$T$ plot in Fig. 6 confirms enhanced low-temperature Hall effect, with a kink at $T_N$ marking the AFM transition. Above 80 K, $|R_H|$ diminishes gradually, demonstrating its high electron density and low carrier mobility.

The Hall resistance of KMn$_6$Bi$_5$ is highly similar to that of RbMn$_6$Bi$_5$ and CsMn$_6$Bi$_5$.[27] This provides further evidence for the distinct behavior of NaMn$_6$Bi$_5$ and the impact of ionic radius variations on the electronic structure. The Hall resistance of RbMn$_6$Bi$_5$ and CsMn$_6$Bi$_5$ deviates from linearity at low temperatures and high fields strongly implies multi-carrier effects, similar to that of KMn$_6$Bi$_5$. This shows that, in addition to ionic radius, temperature can also modulate the charge carrier type in the $A$Mn$_6$Bi$_5$ family.

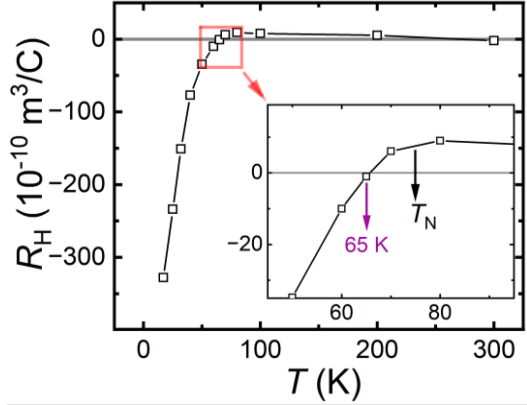

**Fig. 6.** Temperature dependence of the Hall coefficient ($R_H$) for KMn$_6$Bi$_5$ sample A (17–300 K) derived from single-band model fitting. Black arrow indicates the AFM transition at 75 K.

***Conclusions.*** Our systematic magnetotransport study of KMn$_6$Bi$_5$ has revealed a series of distinct, temperature-driven electronic states. At low temperatures (below 5 K), the system enters a correlated two-carrier state, giving rise to a large, non-saturating MR following the $B^{1.8}$ scaling law. A key finding is an emergent quasi-isotropic Lorentz-force response, where the charge carrier dynamics in this fundamentally anisotropic material are exceptionally well-described by an isotropic two-band model. Upon warming, a transport crossover to a single-band, electron-dominated regime occurs around 20 K. This transition is driven by the suppression of the hole pocket's contribution to conduction, likely due to the dramatically reduced hole mobility. Furthermore, we clarified the role of the AFM order. Below $T_N \approx$ 75 K, the onset of AFM phase upon cooling reconstructs the electronic structure, resulting in a partially gapped Fermi surface that leads to a significant reduction in carrier concentration. Within the same temperature range (25-70 K), the suppression of MR by a strong magnetic field indicates that AFM spin fluctuations are a key scattering mechanism. Finally, we demonstrate that the observed low-temperature resistivity upturn under high magnetic fields is not a true phase transition but an intrinsic artifact of MR scaling. These findings establish KMn$_6$Bi$_5$ as a model system for exploring the complex interplay between dimensionality, magnetism, and correlated electron phenomena.

**Acknowledgement**

This work was supported by the National Natural Science Foundation of China (Grant Nos. 12204298 and 12374116), Beijing National Laboratory for Condensed Matter Physics (Grant No. 2023BNLCMPKF019) and the HZNU scientific research and innovation team project (No. TD2025013).